\documentclass[english,aps,preprint, floatfix]{revtex4}
\usepackage[T1]{fontenc}
\usepackage[latin9]{inputenc}
\usepackage{textcomp}
\usepackage{graphicx}

\makeatletter

\providecommand{\tabularnewline}{\\}

\@ifundefined{textcolor}{}
{%
 \definecolor{BLACK}{gray}{0}
 \definecolor{WHITE}{gray}{1}
 \definecolor{RED}{rgb}{1,0,0}
 \definecolor{GREEN}{rgb}{0,1,0}
 \definecolor{BLUE}{rgb}{0,0,1}
 \definecolor{CYAN}{cmyk}{1,0,0,0}
 \definecolor{MAGENTA}{cmyk}{0,1,0,0}
 \definecolor{YELLOW}{cmyk}{0,0,1,0}
}

\makeatother

\usepackage{babel}
\begin{document}

\title{Search for anomalous $WW\gamma$ and $WWZ$ couplings with polarized
$e$-beam at the LHeC}

\author{I.T. Cakir}

\email{ilkayturkcakir@aydin.edu.tr}

\affiliation{Istanbul Aydin University, Faculty of Engineering, Department of
Electrical and Electronics Engineering, 34295, Sefakoy, Istanbul,
Turkey}

\author{O. Cakir}

\email{ocakir@science.ankara.edu.tr}

\affiliation{Ankara University, Faculty of Sciences, Department of Physics, 06100,
Tandogan, Ankara, Turkey}

\author{A. Senol}

\email{asenol@kastamonu.edu.tr}

\affiliation{Kastamonu University, Faculty of Science and Arts, Department of
Physics, 37100, Kuzeykent, Kastamonu, Turkey}

\author{A.T. Tasci}

\email{atasci@kastamonu.edu.tr}

\affiliation{Kastamonu University, Faculty of Science and Arts, Department of
Physics, 37100, Kuzeykent, Kastamonu, Turkey}
\begin{abstract}
We examine the potential of the $ep\to\nu_{e}q\gamma X$ and $ep\to\nu_{e}qZX$
processes to search for the anomalous couplings of $WW\gamma$ and
$WWZ$ vertices at the LHeC with the electron beam energy of $E_{e}=60$
GeV and $E_{e}=140$ GeV. The difference of maximum and minimum bounds
on the anomalous couplings, ($\delta\Delta\kappa_{\gamma}$,$\delta\lambda_{\gamma}$)
and ($\delta\Delta\kappa_{Z}$,$\delta\lambda_{Z}$) are obtained
as{\normalsize{} ($0.990$, $0.122$) and ($0.362$, $0.012$) }without
electron beam polarization at the beam energy of $E_{e}=140$ GeV
for an integrated luminosity of $L_{int}=100$ fb$^{-1}$, respectively.
With the possibility of $e-$beam polarization, we obtain more improved
results as {\normalsize{}($0.975$, $0.118$) and ($0.285$, $0.009$)}
for ($\delta\Delta\kappa_{\gamma},\delta\lambda_{\gamma}$) and ($\delta\Delta\kappa_{Z},\delta\lambda_{Z}$),
respectively. The results are found to be comparable with the current
experimental limits obtained from two-parameter analysis at the lepton
and hadron colliders. It is found that the limits on the anomalous
couplings ($\Delta\kappa_{Z}$,$\lambda_{Z}$) through the process
$ep\to\nu_{e}qZX$ at the LHeC can further improve the current experimental
limits.
\end{abstract}
\maketitle

\section{Introduction}

Triple gauge boson interactions are the consequence of the $SU(2)\times U(1)$
gauge symmetry of the standard model (SM). A precise determination
of the trilinear gauge boson couplings are necessary to test the validity
of the SM and the presence of new physics up to a high energy scale.
Since the tree-level couplings of the $WW\gamma$ and $WWZ$ vertices
are fixed by the SM, any deviations from their SM values would indicate
the new physics beyond the SM. Possible types of the collisions of
the accelerated particles will extend the level of precision to triple
couplings of the gauge bosons. The photoproduction of $W$and $Z$
bosons have been studied in the baseline of lepton-hadron coliders
HERA+LC \cite{Baur89} and the Large Hadron electron Collider (LHeC)
\cite{Brenner12}. An investigation of the potential of the LHeC to
probe anomalous $WW\gamma$ coupling has also been presented in Ref.
\cite{Biswal14}. 

The present bounds on the anomalous $WW\gamma$ and $WWZ$ couplings
are provided by the LEP \cite{LEP13}, Tevatron \cite{CDF10,D012}
and LHC \cite{ATLAS13,CMS13} experiments. Recently, the ATLAS \cite{ATLAS13}
and CMS \cite{CMS13} Collaborations have established updated constraints
on the anomalous $WW\gamma$ and $WWZ$ couplings from the $\gamma W(Z)$
and $W^{+}W^{-}$ production processes. The best available constraints
on $\Delta\kappa_{\gamma}$, $\lambda_{\gamma}$, $\Delta\kappa_{Z}$
and $\lambda_{Z}$ obtained from one parameter analysis at different
experiments are summarized in Table \ref{tab:tab1}. 

The limits based on one-parameter analysis at $95\%$ C.L. on the
$\Delta\kappa_{\gamma}$ and $\lambda_{\gamma}$ from ATLAS Collaboration
with $W\gamma$ production process data at $\sqrt{s}=7$ TeV and $L_{int}=4.6$
fb$^{-1}$ are $(-0.135,0.190)$ and $(-0.065,0.061)$ \cite{ATLAS13}.
Same limits from the $95\%$ C.L. two-parameter analysis are $(-0.420,0.480)$
for $\Delta\kappa_{\gamma}$ and $(-0.068,0.062)$ for $\lambda_{\gamma}$.
Two-parameter $95\%$ C.L. limits on anomalous couplings $\Delta\kappa_{Z}$
and $\lambda_{Z}$ are given as $(-0.045,0.045)$ and $(-0.063,0.063)$,
respectively. %The limits on these anomalous couplings are also found in Ref.
%\cite{ATLAS13} for different scenarios.

According to CMS Collaboration, one-parameter $95\%$ C.L. limits
are $(-0.210,0.220)$ and $(-0.048,0.048)$ for $\Delta\kappa_{\gamma}$
and $\lambda_{\gamma}$ from $W\gamma$ production process at $\sqrt{s}=7$
TeV and $L_{int}=5$ fb$^{-1}$ \cite{CMS13}. From two-parameter
contours, the upper limits for $\Delta\kappa_{\gamma}$ and $\lambda_{\gamma}$
are obtained as $(-0.250,0.250)$ and $(-0.050,0.042)$ at the $95\%$
C.L. while one-parameter $95\%$ C.L. limits on $\Delta\kappa_{Z}$
and $\lambda_{Z}$ are $(-0.160,0.157)$ and $(-0.048,0.048)$ from
$W^{+}W^{-}$ production process at $\sqrt{s}=7$ TeV. Here, the relation
$\Delta\kappa_{Z}=\Delta g_{1}^{Z}-\Delta\kappa_{\gamma}\cdot\tan^{2}\theta_{W}$
is used to extract some of the the limits in the LEP scenario. The
results from ATLAS and CMS experiments based on two parameter analysis
of the anomalous couplings are given in Table \ref{tab:tab2}.

In this study, we examined the $ep\to\nu_{e}q\gamma X$ and $ep\to\nu_{e}qZX$
processes with anomalous $WW\gamma$ and $WWZ$ couplings at the high
energy electron-proton collider namely, the Large Hadron electron
Collider (LHeC). This collider is considered to be realised by accelerating
electrons in a linear accelerator (linac) to 60 \textminus{} 140 GeV
and colliding them with the 7 TeV protons incoming from the LHC. We
take into account the possibility of the electron beam polarization
at LHeC which extends the sensitivity to anomalous triple gauge boson
couplings. The anticipated integrated luminosity is about at the order
of 10 and 100 fb$^{-1}$ \cite{AbelleiraFernandez:2012cc}.

\begin{table}
\protect\caption{The available $95\%$ C.L. bounds on anomalous couplings ($\Delta\kappa_{\gamma}$,$\lambda_{\gamma}$)
and ($\Delta\kappa_{Z}$,$\lambda_{Z}$) from the data at LEP, Tevatron,
and LHC experiments. In each case the parameter listed is varied while
the others are fixed to their SM values. \label{tab:tab1}}

\begin{tabular}{|c|c|c|c|c|c|}
\hline 
 & LEP\cite{LEP13} & CDF\cite{CDF10} & D0\cite{D012} & ATLAS\cite{ATLAS13} & CMS\cite{CMS13}\tabularnewline
\hline 
\hline 
$\Delta\kappa_{\gamma}$ & {[}-0.099, 0.066{]} & {[}-0.460, 0.390{]} & {[}-0.158, 0.255{]} & {[}-0.135, 0.190{]} & {[}-0.210, 0.220{]}\tabularnewline
\hline 
$\lambda_{\gamma}$ & {[}-0.059, 0.017{]} & {[}-0.180, 0.170{]} & {[}-0.036, 0.044{]} & {[}-0.065, 0.061{]} & {[}-0.048, 0.048{]}\tabularnewline
\hline 
$\Delta\kappa_{Z}$ & {[}-0.073, 0.050{]} & {[}-0.414, 0.470{]} & {[}-0.110, 0.131{]} & {[}-0.061, 0.093{]} & {[}-0.160, 0.157{]}\tabularnewline
\hline 
$\lambda_{Z}$ & {[}-0.059, 0.017{]} & {[}-0.140, 0.150{]} & {[}-0.036, 0.044{]} & {[}-0.062, 0.065{]} & {[}-0.048, 0.048{]}\tabularnewline
\hline 
\end{tabular}
\end{table}

\begin{table}
\protect\caption{The available $95\%$ C.L. two-parameter bounds on anomalous couplings
($\Delta\kappa_{\gamma}$,$\lambda_{\gamma}$) and ($\Delta\kappa_{Z}$,$\lambda_{Z}$)
from the ATLAS and CMS experiments. The difference of maximum and
minimum bounds are show in last two column. \label{tab:tab2}}

\begin{tabular}{|c|c|c|c|c|}
\hline 
 & ATLAS\cite{ATLAS13} & CMS\cite{CMS13} & ATLAS (max-min) & CMS (max-min)\tabularnewline
\hline 
\hline 
$\Delta\kappa_{\gamma}$ & {[}-0.420, 0.480{]} & {[}-0.250, 0.250{]} & 0.900 & 0.500\tabularnewline
\hline 
$\lambda_{\gamma}$ & {[}-0.068, 0.062{]} & {[}-0.050, 0.042{]} & 0.130 & 0.092\tabularnewline
\hline 
$\Delta\kappa_{Z}$ & {[}-0.045, 0.045{]} & {[}-0.160, 0.180{]} & 0.090 & 0.340\tabularnewline
\hline 
$\lambda_{Z}$ & {[}-0.063, 0.063{]} & {[}-0.055, 0.055{]} & 0.126 & 0.110\tabularnewline
\hline 
\end{tabular}
\end{table}

\section{Anomalous Couplings}

The $WW\gamma$ and $WWZ$ interaction vertices are described by an
effective Lagrangian with the coupling constants $g_{WW\gamma}$ and
$g_{WWZ}$ and dimensionless parameter pairs ($\kappa_{\gamma},\lambda_{\gamma}$)
and ($\kappa_{Z},\lambda_{Z}$), 

\begin{eqnarray}
{\cal L} & = & ig_{WW\gamma}[g_{1}^{\gamma}(W_{\mu\nu}^{\dagger}W^{\mu}A^{\nu}-W^{\mu\nu}W_{\mu}^{\dagger}A_{\nu})+\kappa_{\gamma}W_{\mu}^{\dagger}W_{\nu}A^{\mu\nu}+\frac{\lambda_{\gamma}}{m_{W}^{2}}W_{\rho\mu}^{\dagger}W_{\nu}^{\mu}A^{\nu\rho}]\nonumber \\
 & + & ig_{WWZ}[g_{1}^{Z}(W_{\mu\nu}^{\dagger}W^{\mu}Z^{\nu}-W^{\mu\nu}W_{\mu}^{\dagger}Z_{\nu})+\kappa_{Z}W_{\mu}^{\dagger}W_{\nu}Z^{\mu\nu}+\frac{\lambda_{Z}}{m_{W}^{2}}W_{\rho\mu}^{\dagger}W_{\nu}^{\mu}Z^{\nu\rho}]\label{eq:eq1}
\end{eqnarray}
where $g_{WW\gamma}=g_{e}=gsin\theta_{W}$ and $g_{WWZ}=gcos\theta_{W}$.
In general these vertices involve six C and P conserving couplings
\cite{Hagivara87}. However, the electromagnetic gauge invariance
requires that $g_{1}^{\gamma}=1$. The anomalous couplings are defined
as $\kappa_{V}=1+\triangle\kappa_{V}$ where $V=\gamma,\, Z$ and
$g_{1}^{Z}=1+\triangle g_{1}^{Z}.$ The $W_{\mu\nu}$, $Z_{\mu\nu}$
and $A_{\mu\nu}$ are the field strength tensors for the $W$- boson,
$Z-$boson and photon, respectively.

The values of the couplings $\kappa_{\gamma}=\kappa_{Z}=1$ and $\lambda_{\gamma}=\lambda_{Z}=0$
correspond to the case of the SM. Since unitarity restricts the $WW\gamma$
and $WWZ$ couplings to their SM values at very high energies, the
triple gauge couplings are modified as $\Delta\kappa_{V}(q^{2})=\Delta\kappa_{V}(0)/(1+q^{2}/\Lambda^{2})^{2}$
and $\lambda_{V}(q^{2})=\lambda_{V}(0)/(1+q^{2}/\Lambda^{2})^{2}$
where $V=\gamma$,$Z$. The $q^{2}$ is the square of momentum transfer
into the process and $\Lambda$ is the new physics energy scale. The
$\Delta\kappa_{V}(0)$ and $\lambda_{V}(0)$ are the values of the
anomalous couplings at $q^{2}=0$. We assume the values of the anomalous
couplings remain approximate constant in the interested energy scale.
We have implemented interactions terms in the CalcHEP \cite{CalcHEP7}.

\begin{figure}
\includegraphics{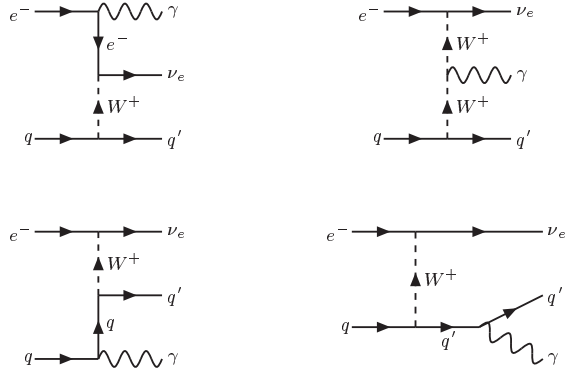}

\protect\caption{Representative Feynman diagrams for subprocesses $eq\rightarrow\nu\gamma q$.
\label{fig:fig1}}
\end{figure}

\begin{figure}
\includegraphics{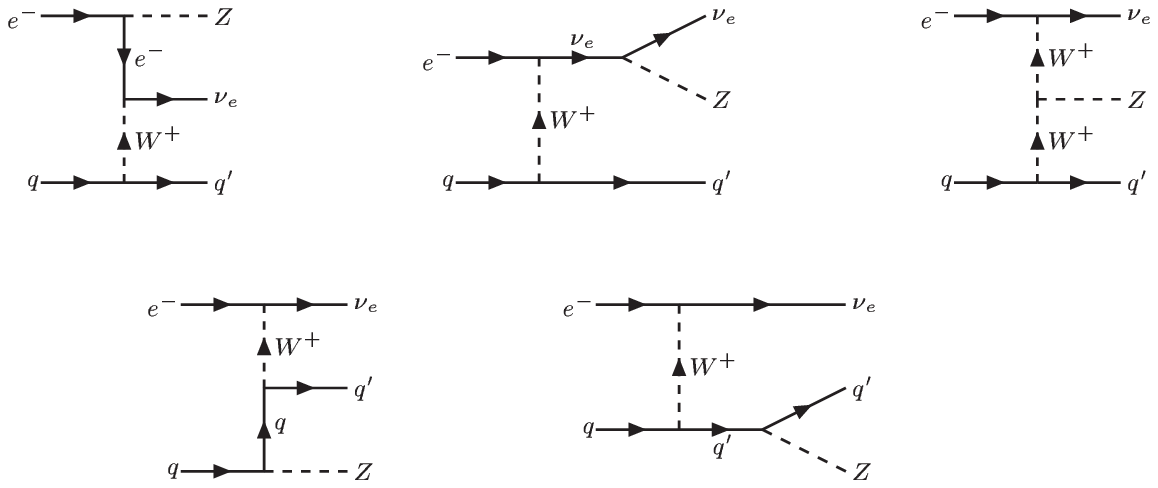}

\protect\caption{Representative Feynman diagrams for subprocesses $eq\rightarrow\nu Zq$.
\label{fig:fig2}}
\end{figure}

\section{Production Cross Sections}

\begin{figure}
\includegraphics{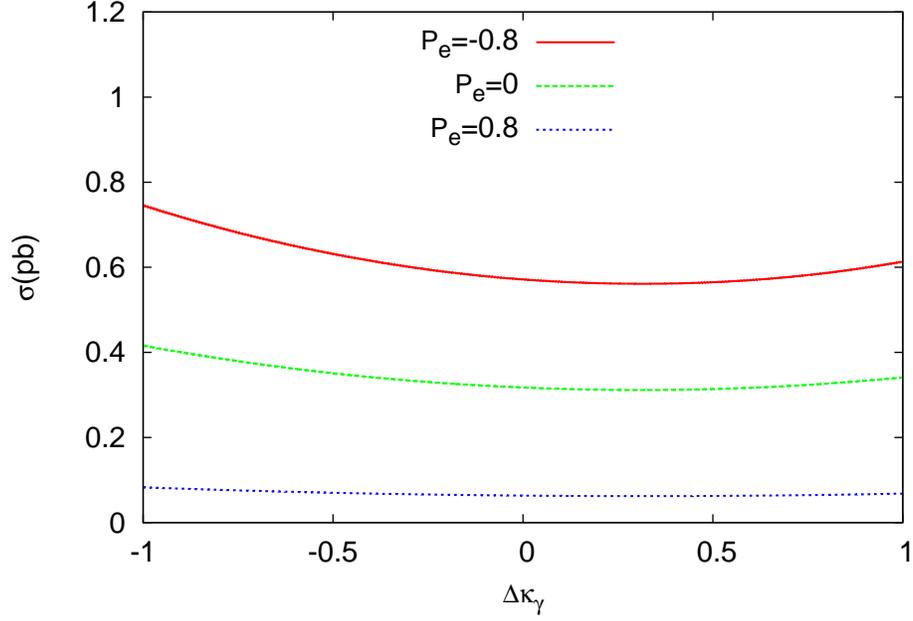}

\protect\caption{The cross section depending on anomalous coupling $\Delta\kappa_{\gamma}$
of the process $ep\rightarrow\nu\gamma qX$ for $E_{e}=60$ GeV. \label{fig:fig3}}
\end{figure}

\begin{figure}
\includegraphics{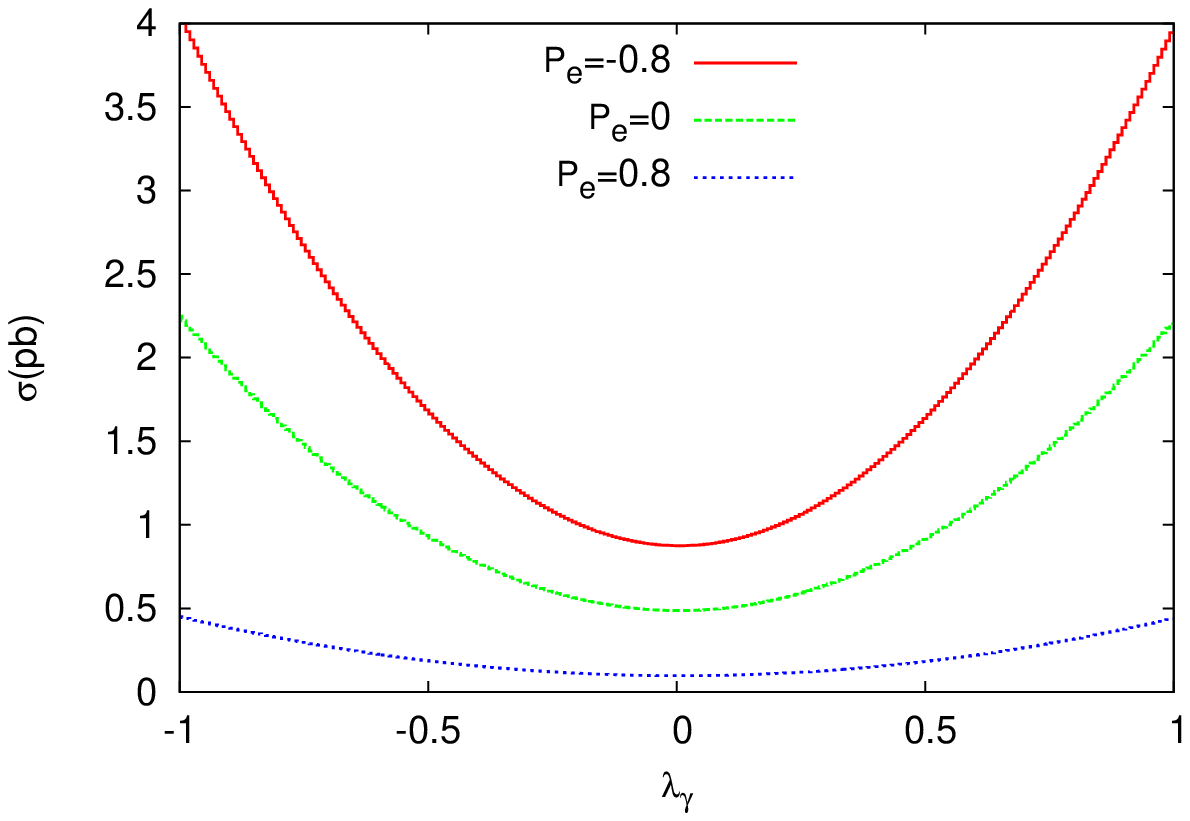}

\protect\caption{The cross section depending on anomalous coupling $\lambda_{\gamma}$
of the process $ep\rightarrow\nu\gamma qX$ for $E_{e}=60$ GeV. \label{fig:fig4}}
\end{figure}

\begin{figure}
\includegraphics{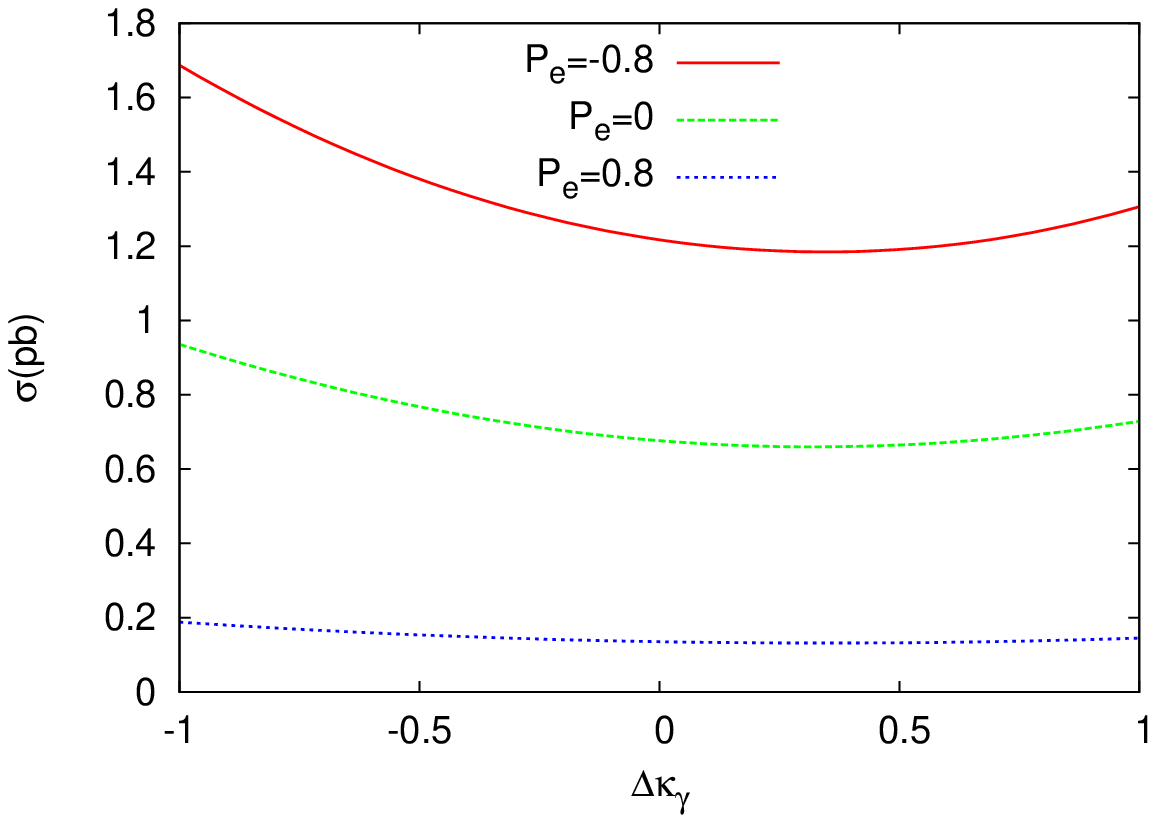}

\protect\caption{The same as Fig 3 but for $E_{e}=140$ GeV. \label{fig:fig5}}
\end{figure}

\begin{figure}
\includegraphics{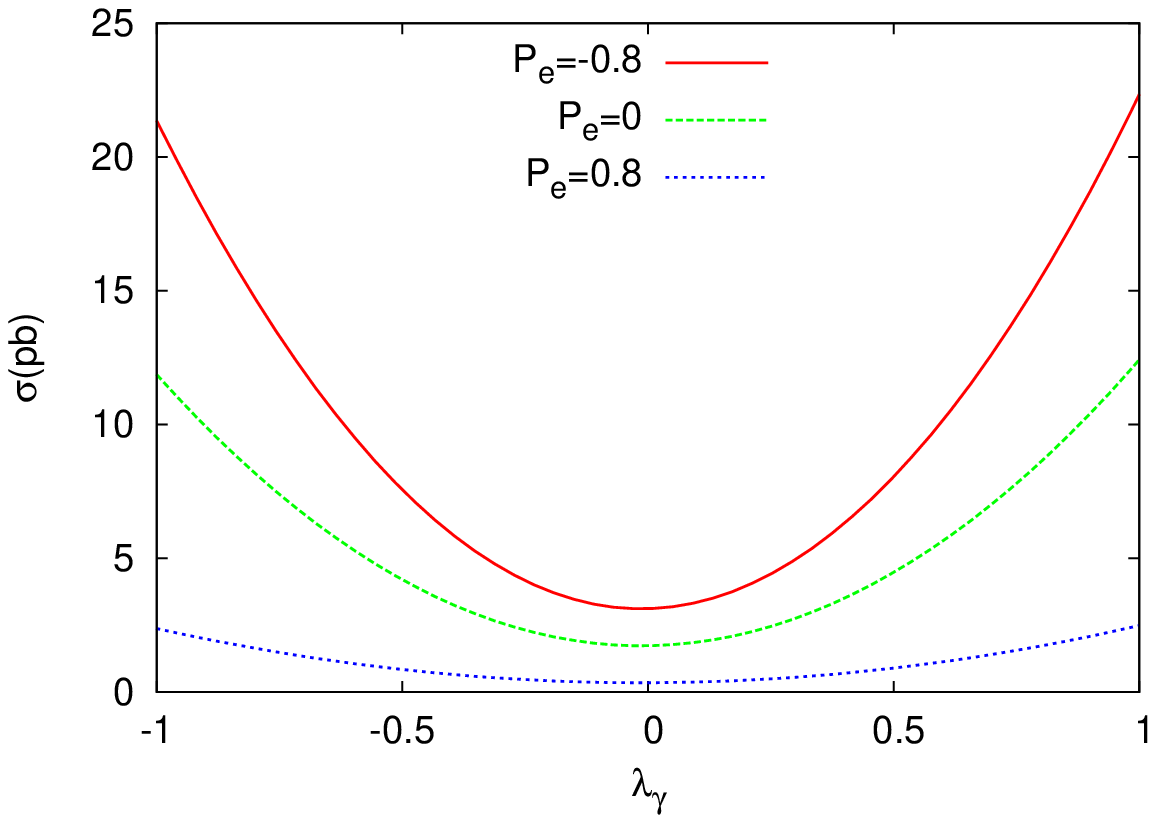}

\protect\caption{The same as Fig 4 but for $E_{e}=140$ GeV.\label{fig:fig6}}
\end{figure}

According to the effective Lagrangian the anomalous vertices for triple
gauge interactions $WW\gamma$ and $WWZ$ are presented in the Feynman
graphs as shown in Fig. \ref{fig:fig1} and Fig. \ref{fig:fig2}.
In order to calculate the cross sections for the process $ep\to\nu_{e}q\gamma X$
and $ep\to\nu_{e}qZX$, we apply the transverse momentum cut on photon
and jet as $p_{T}^{\gamma}>50$ GeV, $p_{T}^{j}>20$ GeV; missing
transverse momentum cut $p_{T}^{\nu}>20$ GeV, pseudorapidity cuts
$|\eta_{\gamma,j}|<3.5$; a cone radius cut between photons and jets
$\Delta R_{\gamma,j}>1.5$. Using these cuts and CTEQ6L \cite{Pumplin:2002vw}
for parton distribution functions, the total cross sections of the
process $ep\rightarrow\nu\gamma qX$ as a function of anomalous couplings
$\Delta\kappa_{\gamma}$ and $\lambda_{\gamma}$ for $E_{e}=60$ GeV
($140$ GeV) with ($P_{e}=\pm0.8$) and without ($P_{e}=0$) electron
beam polarization are presented in Figs. \ref{fig:fig3} and \ref{fig:fig4}
(Figs. \ref{fig:fig5} and \ref{fig:fig6}), respectively. It is clear
from these figures that the polarization ($P_{e}=-0.8$) enhances
the cross sections according to the unpolarized case. 

\begin{figure}
\includegraphics{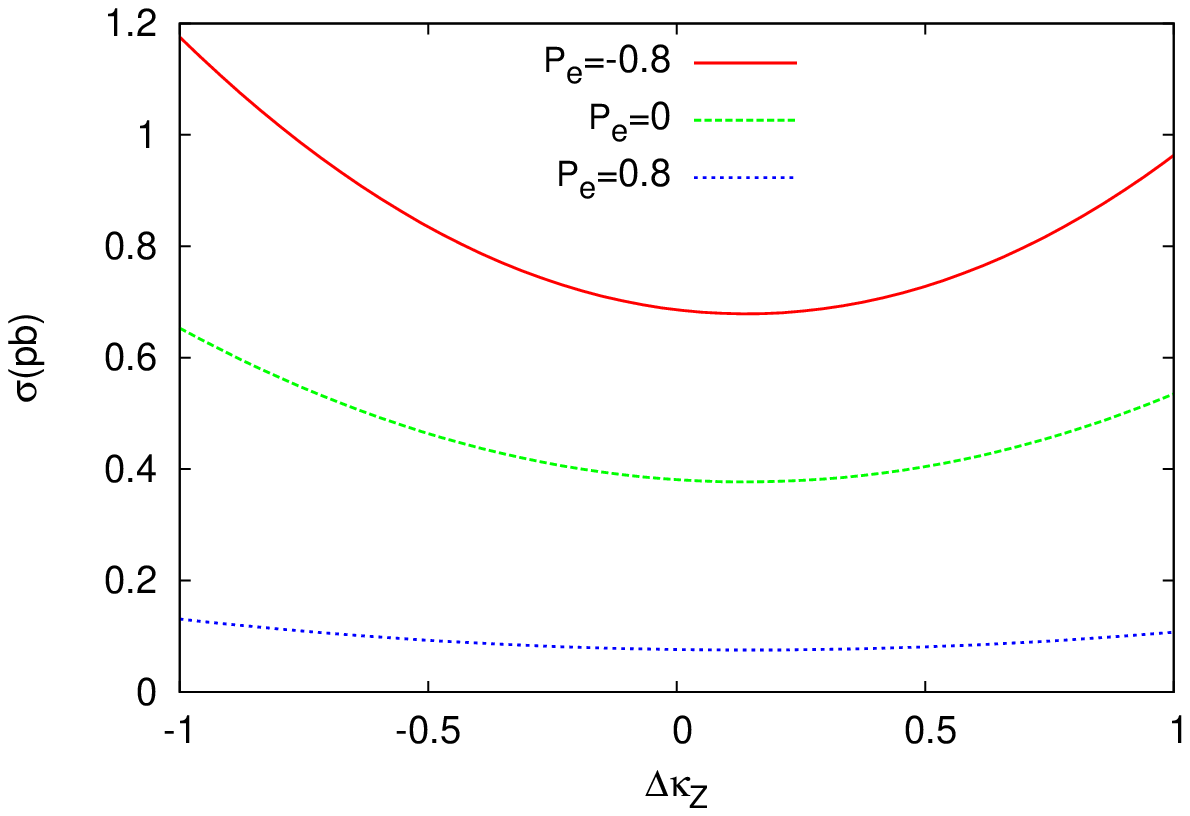}

\protect\caption{The cross section depending on anomalous $\Delta\kappa_{z}$ coupling
of the process $ep\rightarrow\nu ZqX$ for $E_{e}=60$ GeV. \label{fig:fig7}}
\end{figure}

\begin{figure}
\includegraphics{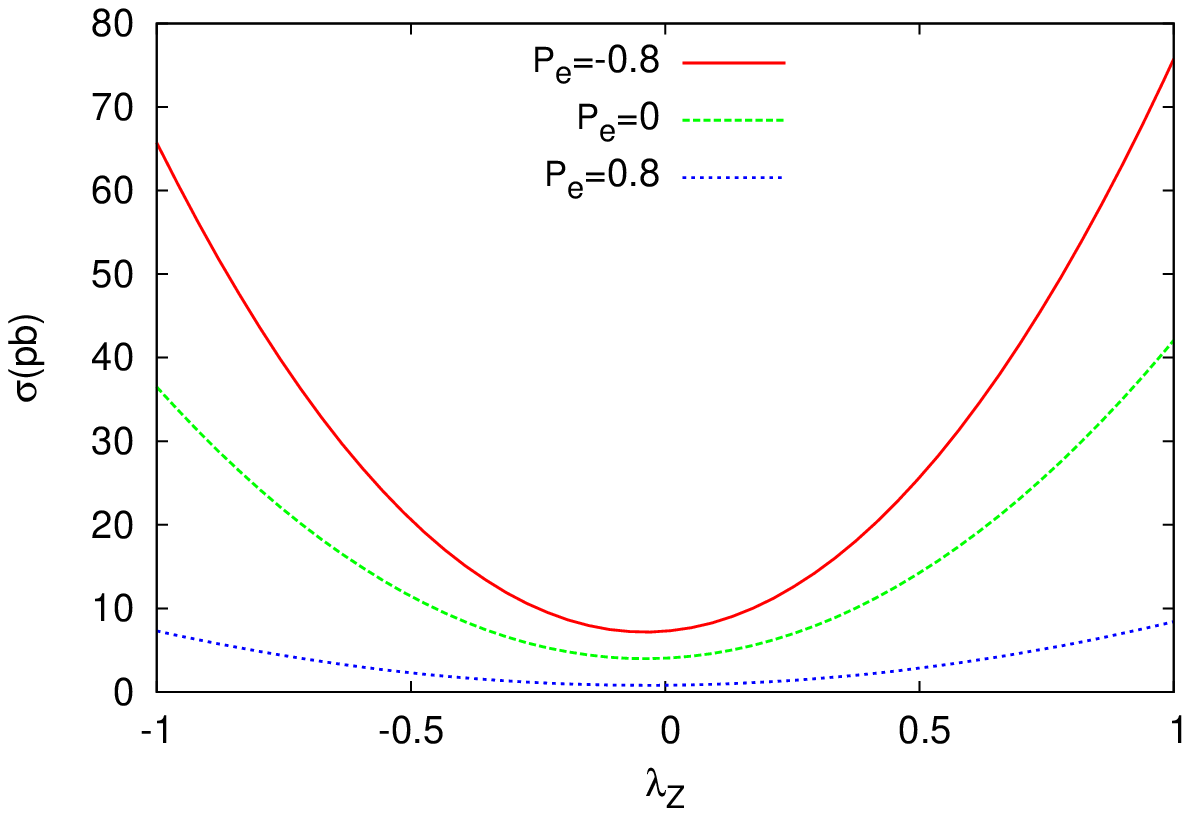}

\protect\caption{The cross section depending on anomalous $\lambda_{z}$ coupling of
the process $ep\rightarrow\nu ZqX$ for $E_{e}=60$ GeV. \label{fig:fig8}.}
\end{figure}

\begin{figure}
\includegraphics{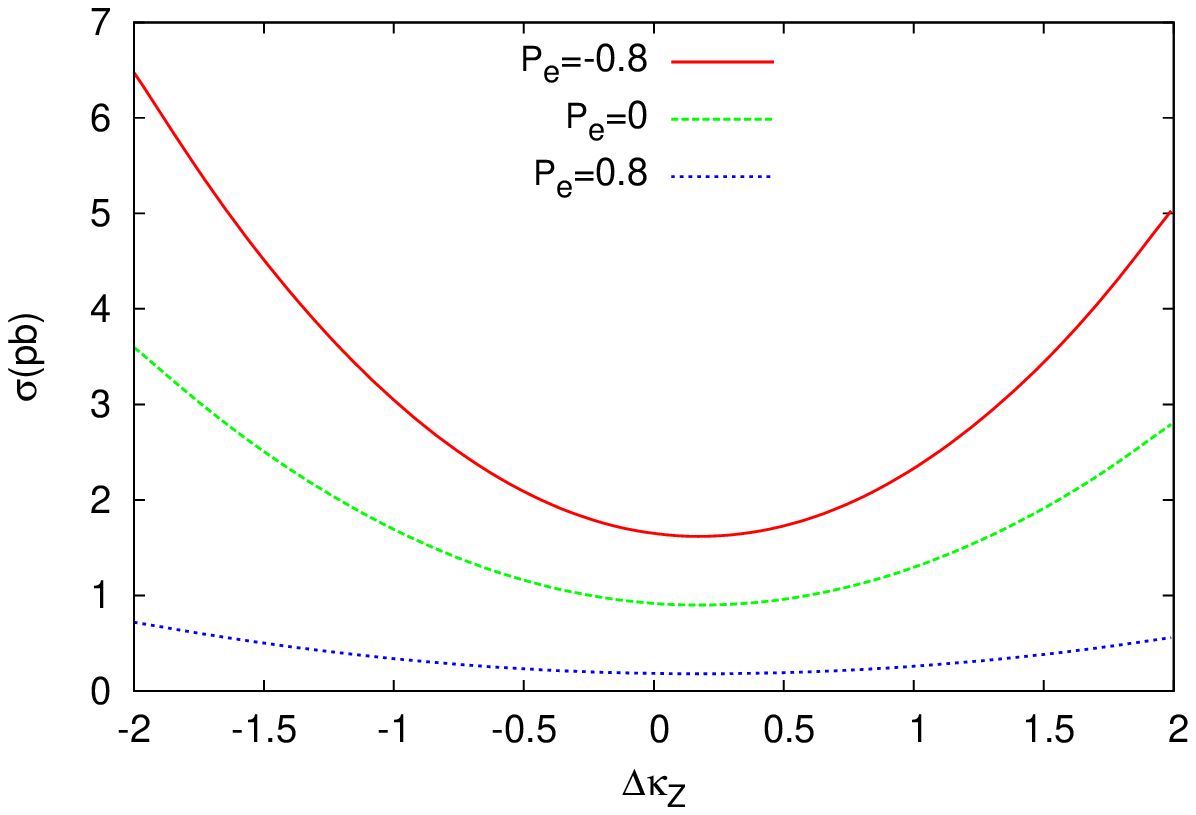}

\protect\caption{The same as Fig 7 but for $E_{e}=140$ GeV. \label{fig:fig9}}
\end{figure}

\begin{figure}
\includegraphics{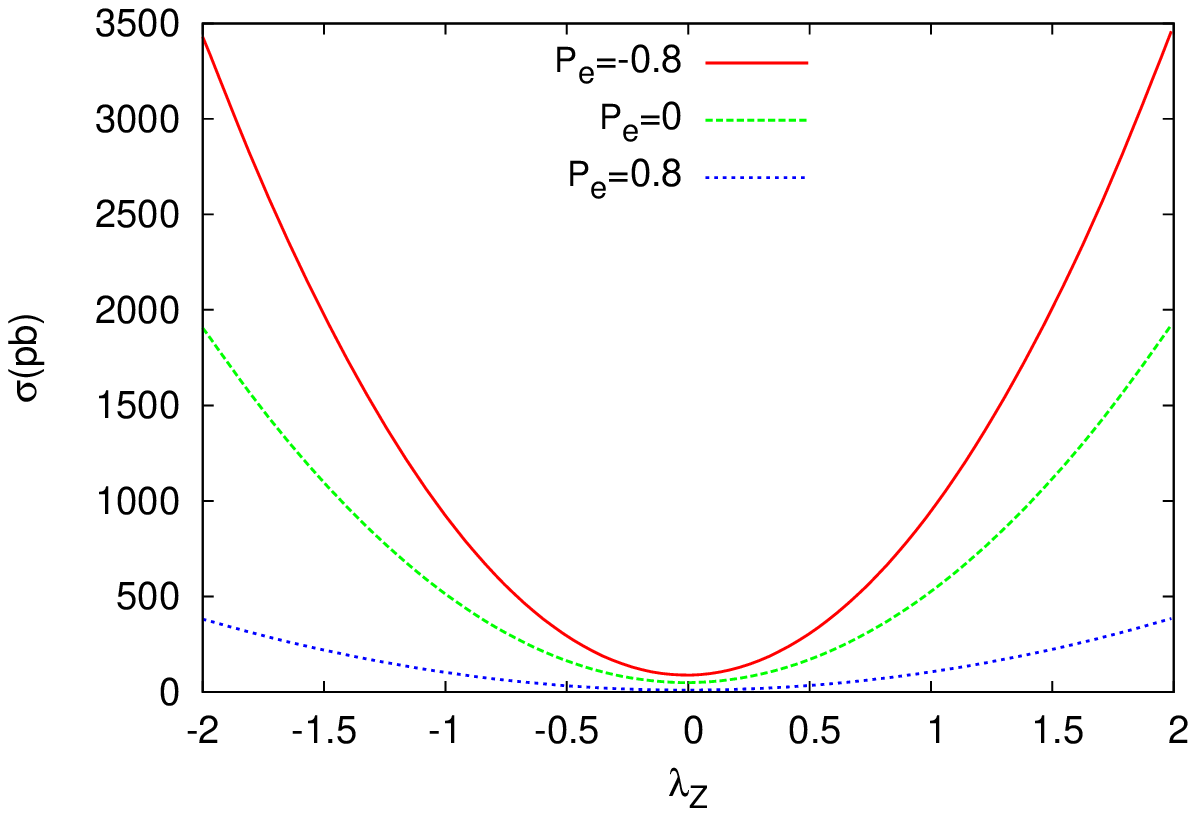}

\protect\caption{The same as Fig 8 but for $E_{e}=140$ GeV. \label{fig:fig10}.}
\end{figure}

The cross sections depending on anomalous couplings $\Delta\kappa_{Z}$
and $\lambda_{Z}$ of the process $ep\rightarrow\nu ZqX$ for $E_{e}=60$
GeV ($140$ GeV) with ($P_{e}=\pm0.8$) and without ($P_{e}=0$) electron
beam polarization are presented in Figs. \ref{fig:fig7} and \ref{fig:fig8}
(Figs. \ref{fig:fig9} and \ref{fig:fig10}), respectively.

\section{Analysis}

\begin{figure}
\includegraphics{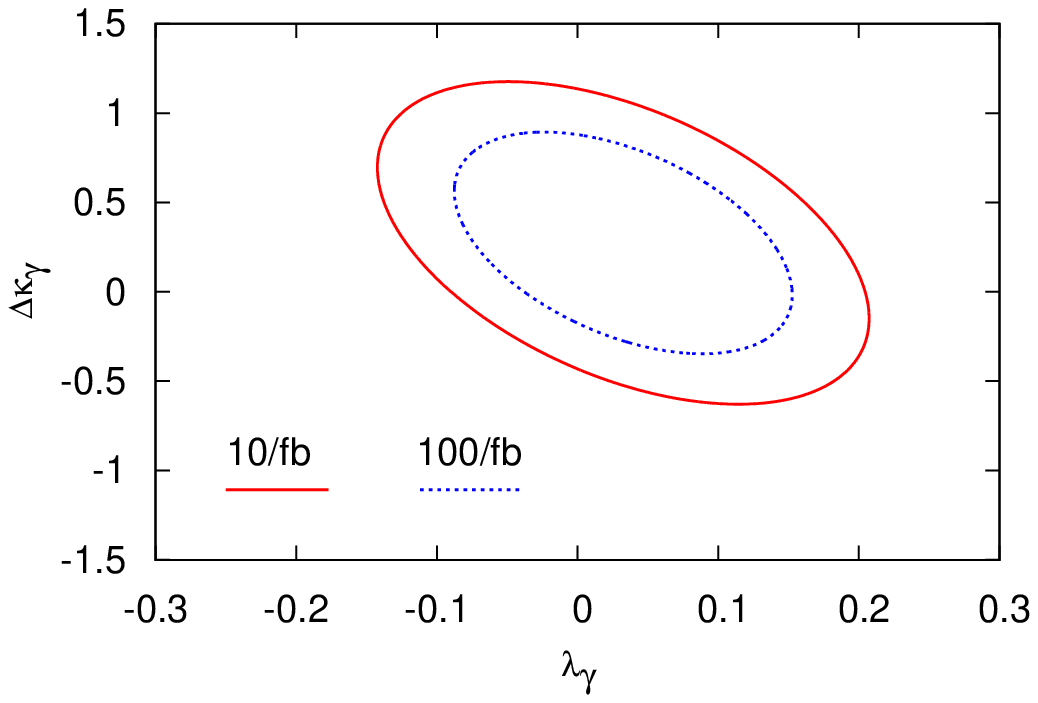}

\protect\caption{Contour plots in the $\lambda_{\gamma}-\Delta\kappa_{\gamma}$ plane
for the integrated luminosity of $10$ fb$^{-1}$ and $100$ fb$^{-1}$
at electron beam energy $E_{e}=60$ GeV with the beam polarization
$P_{e}=0.8$. \label{fig:fig11}}
\end{figure}

\begin{figure}
\includegraphics{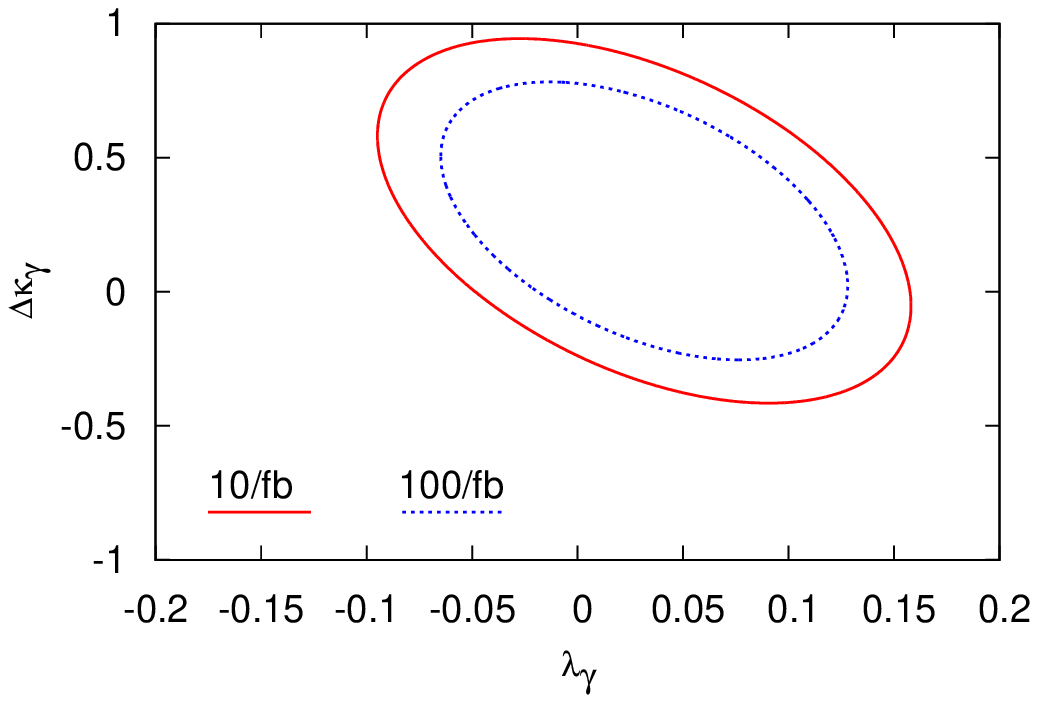}

\protect\caption{The same as Fig. 11, but for $P_{e}=0$. \label{fig:fig12}}
\end{figure}

\begin{figure}
\includegraphics{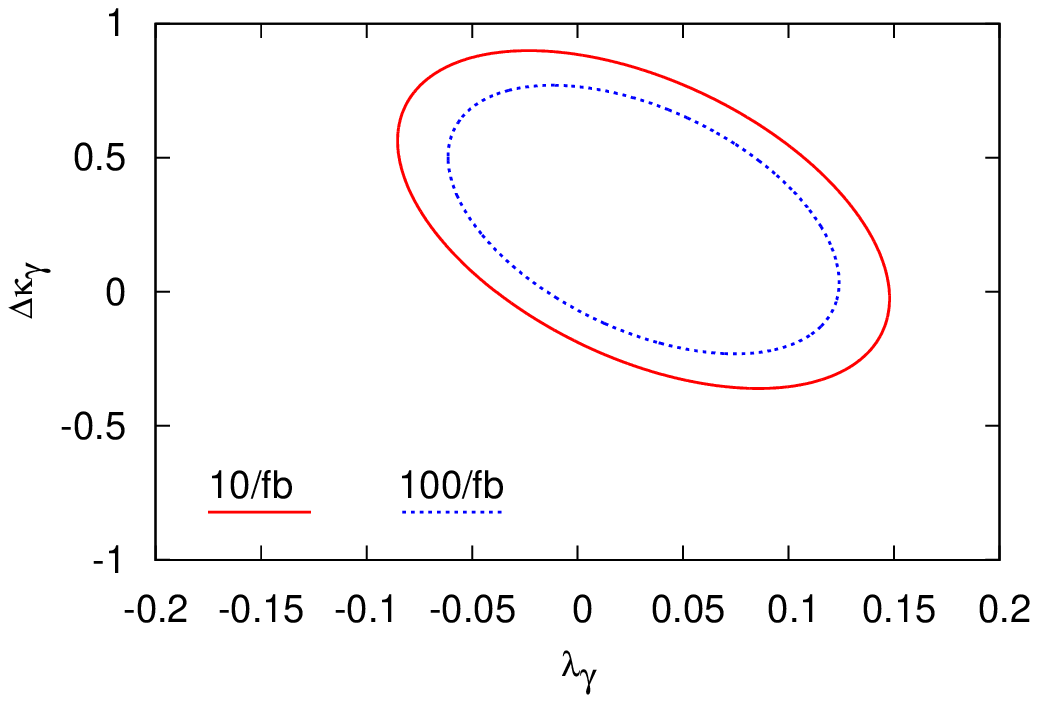}

\protect\caption{The same as Fig. 11, but for $P_{e}=-0.8$. \label{fig:fig13}}
\end{figure}

\begin{figure}
\includegraphics{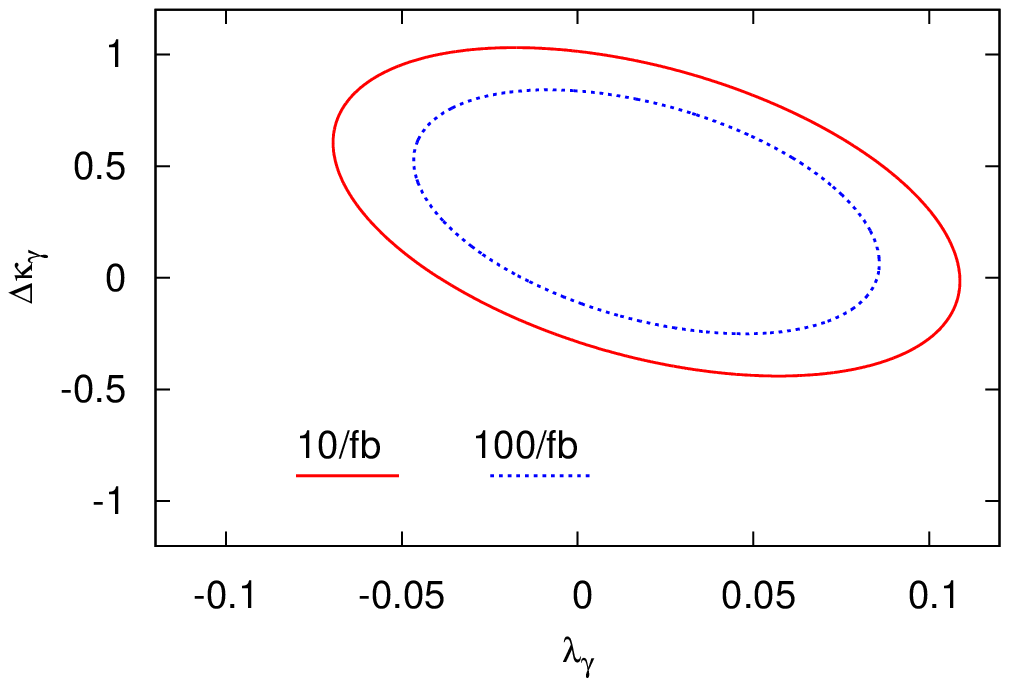}

\protect\caption{Contour plots in the $\lambda_{\gamma}-\Delta\kappa_{\gamma}$ plane
for the integrated luminosity of $10$ fb$^{-1}$ and $100$ fb$^{-1}$
at electron beam energy $E_{e}=140$ GeV with polarization $P_{e}=0.8$.
\label{fig:fig14}}
\end{figure}

\begin{figure}
\includegraphics{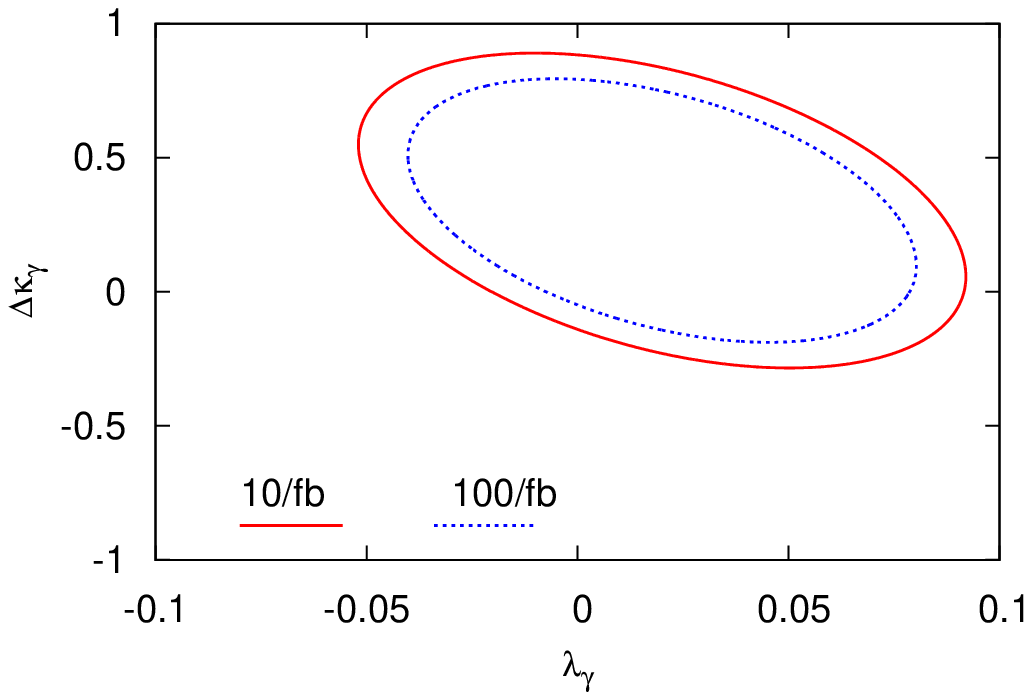}

\protect\caption{The same as Fig. 14, but for $P_{e}=0$. \label{fig:fig15}}
\end{figure}

\begin{figure}
\includegraphics{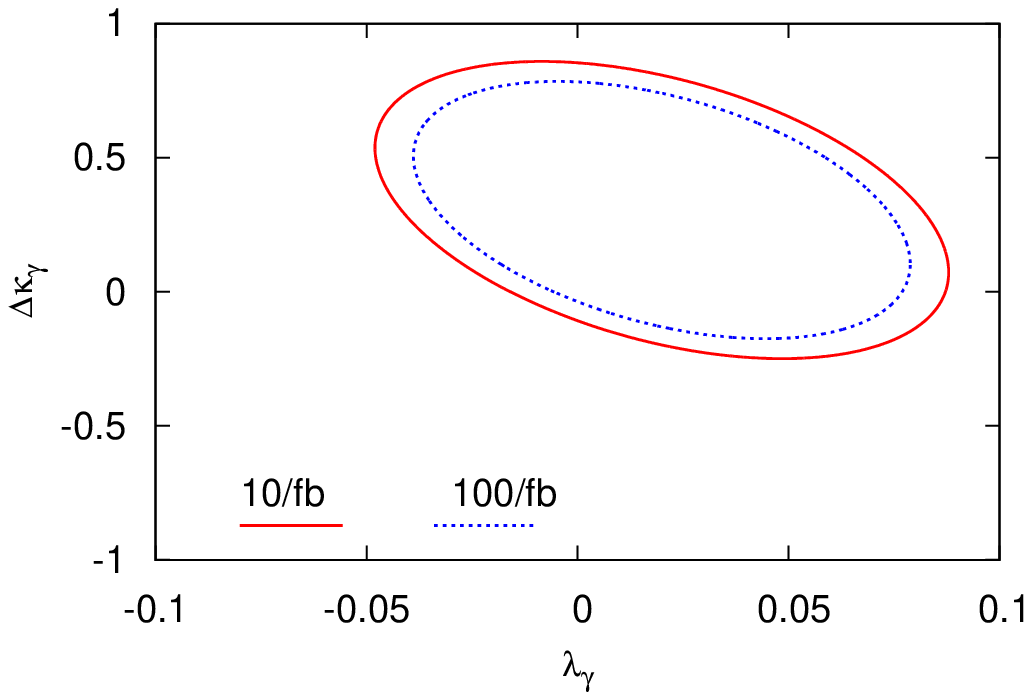}

\protect\caption{The same as Fig. 14, but for $P_{e}=-0.8$ GeV. \label{fig:fig16}}
\end{figure}

\begin{figure}
\includegraphics{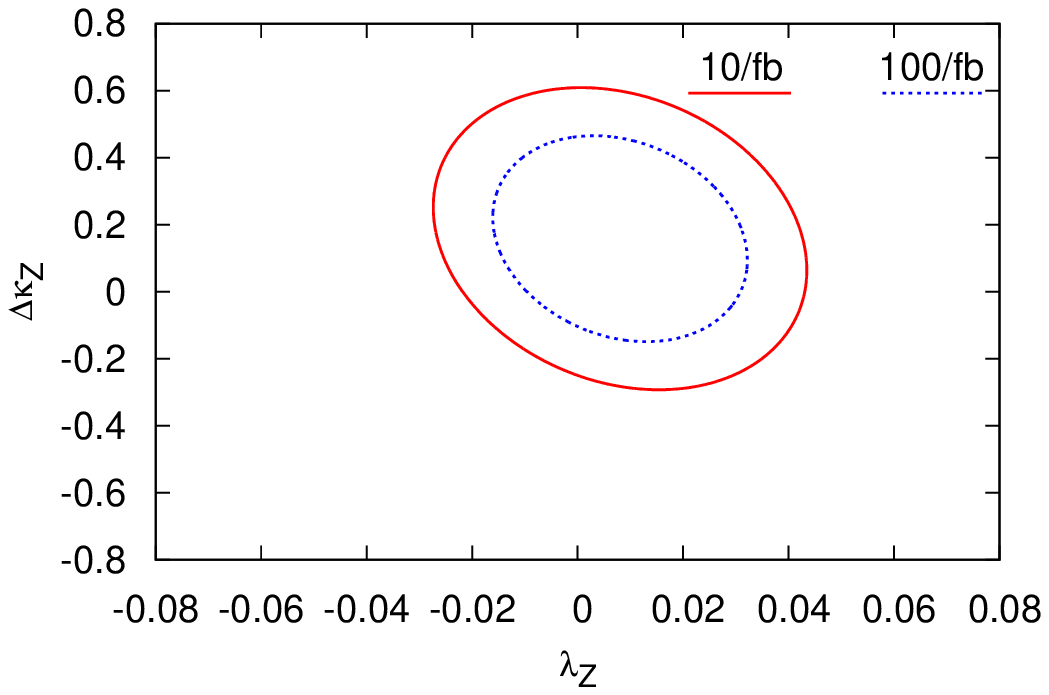}

\protect\caption{Contour plots in the $\lambda_{z}-\Delta\kappa_{z}$ plane for the
integrated luminosity of $10$ fb$^{-1}$ and $100$ fb$^{-1}$ at
electron beam energy $E_{e}=60$ GeV with polarization $P_{e}=0.8$.
\label{fig:fig17}}
\end{figure}

\begin{figure}
\includegraphics{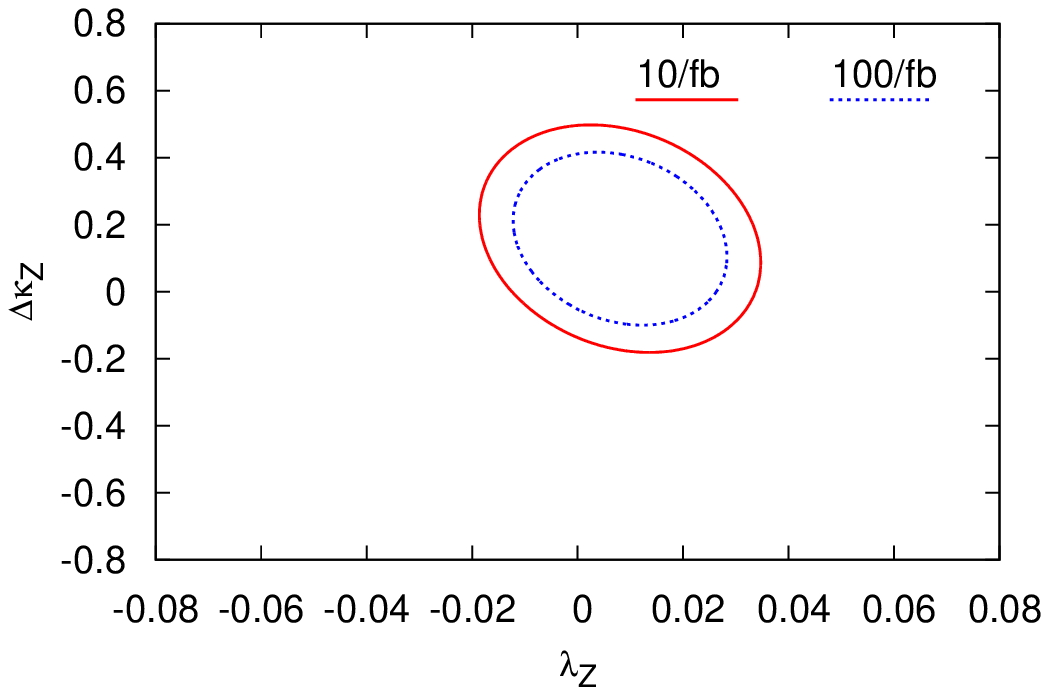}

\protect\caption{The same as Fig. 17, but for $P_{e}=0$. \label{fig:fig18}}
\end{figure}

\begin{figure}
\includegraphics{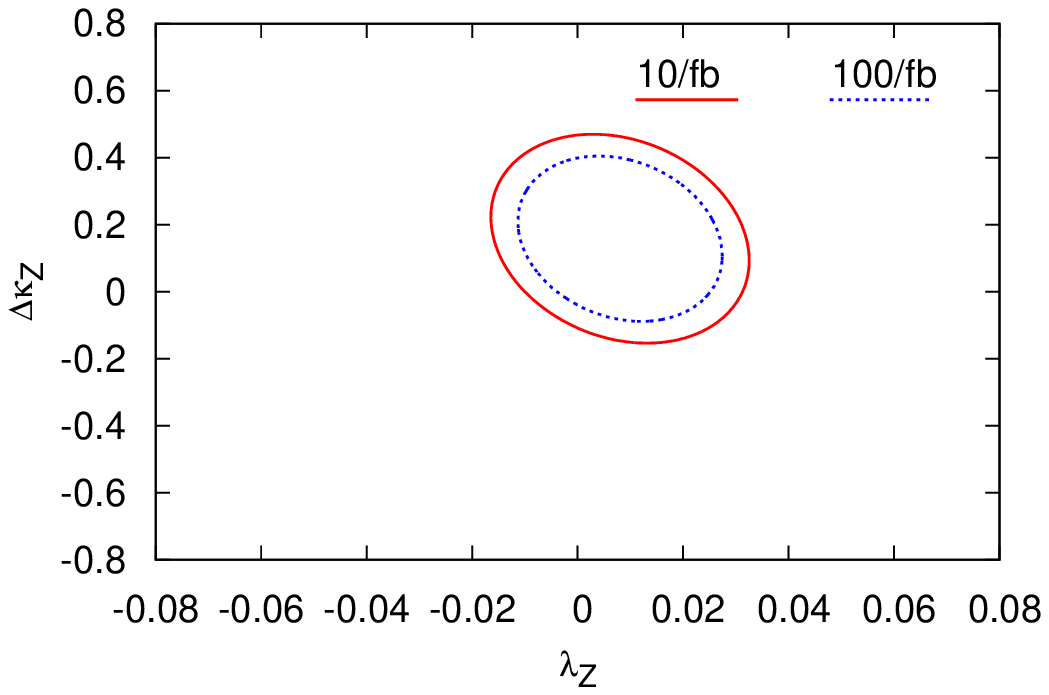}

\protect\caption{The same as Fig. 17, but for $P_{e}=-0.8$. \label{fig:fig19}}
\end{figure}

\begin{figure}
\includegraphics{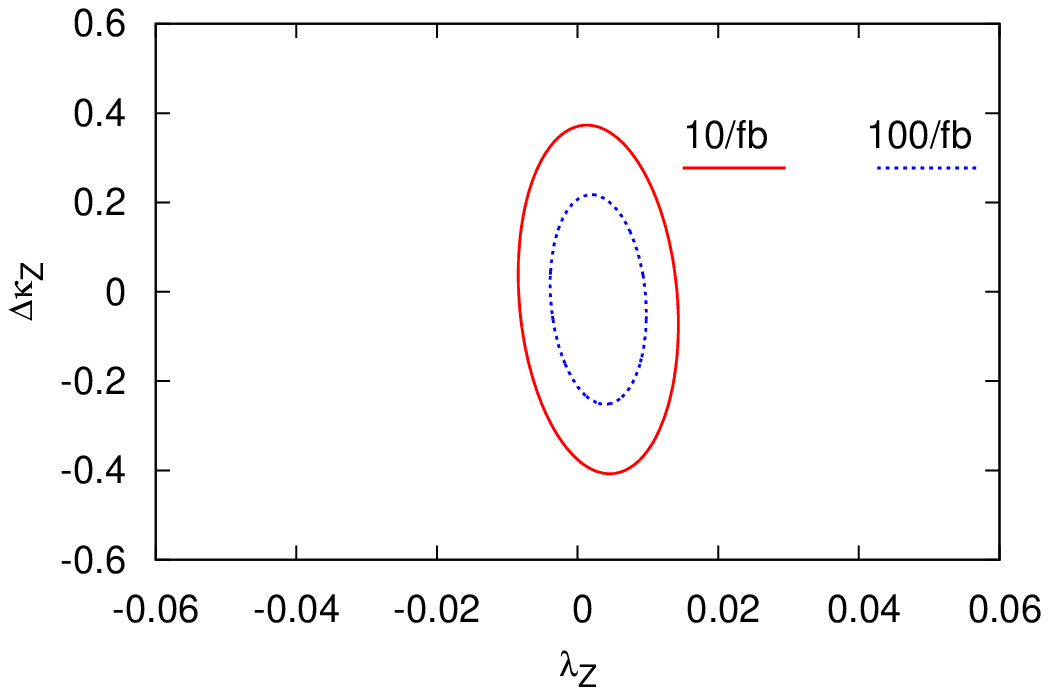}

\protect\caption{Contour plots in the $\lambda_{z}-\Delta\kappa_{z}$ plane for the
integrated luminosity of $10$ fb$^{-1}$ and $100$ fb$^{-1}$ at
electron beam energy $E_{e}=140$ GeV with polarization $P_{e}=0.8$.
\label{fig:fig20}}
\end{figure}

\begin{figure}
\includegraphics{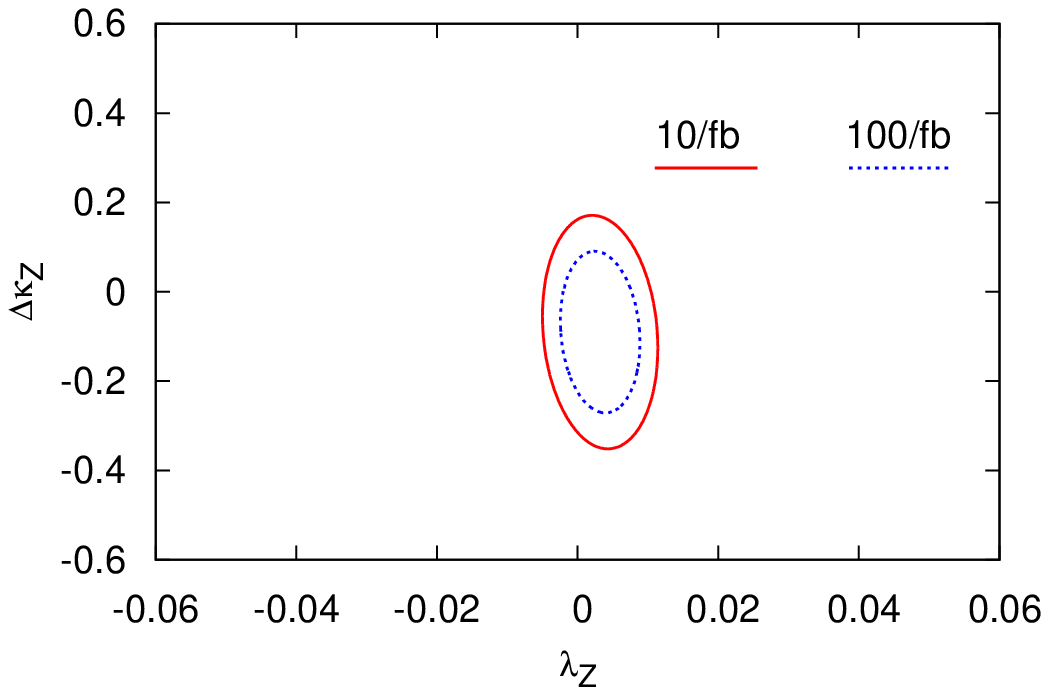}

\protect\caption{The same as Fig. 20, but for $P_{e}=0$. \label{fig:fig21}}
\end{figure}

\begin{figure}
\includegraphics{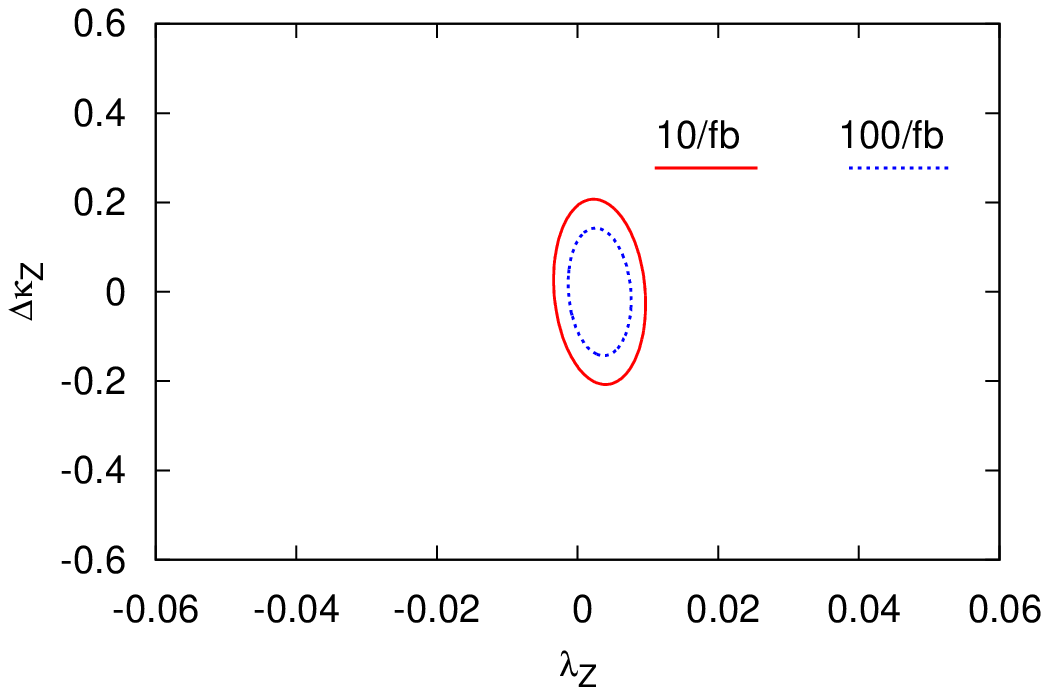}

\protect\caption{The same as Fig. 20, but for $P_{e}=-0.8$. \label{fig:fig22}}
\end{figure}

In order to estimate the sensitivity to the anomalous $WW\gamma$
and $WWZ$ couplings, we use two-parameter $\chi^{2}$ function:

\begin{equation}
\chi^{2}(\Delta\kappa_{V},\lambda_{V})=\left(\frac{\sigma_{SM}-\sigma(\Delta\kappa_{V},\lambda_{V})}{\Delta\sigma_{SM}}\right)^{2}\label{eq:eq.2}
\end{equation}
where $\Delta\sigma_{SM}=\sigma_{SM}\sqrt{\delta_{stat.}^{2}}$ with
$\delta_{stat.}=1/\sqrt{N_{SM}}$ and $N_{SM}=\sigma_{SM}\cdot\epsilon\cdot L_{int}$.
In our calculations, we consider that two of the couplings ($\Delta\kappa$,$\lambda$)
are assumed to deviate from their SM value. We estimate the sensitivity
to the anomalous couplings at LHeC for the integrated luminosities
of $10$ and $100$ fb$^{-1}$. The contour plots in $\triangle\kappa_{\gamma}-\lambda_{\gamma}$
plane for the integrated luminosities of 10 fb$^{-1}$ and 100 fb$^{-1}$
at electron beam energies $E_{e}=60$ ($140$) GeV with different
polarizations are given in Figs. \ref{fig:fig11}-\ref{fig:fig16}.
For the process $ep\to\nu_{e}qZX$, we make analysis of the signal
and backgrounds when $Z$ decays into leptonically, $Z\rightarrow l^{+}l^{-}$
where $l=e,$$\mu$. The contour plots in $\triangle\kappa_{Z}-\lambda_{Z}$
plane for the integrated luminosities of 10 fb$^{-1}$ and 100 fb$^{-1}$
at electron beam energies E=60 (140) GeV with different polarizatons
are presented in Figs. \ref{fig:fig17}-\ref{fig:fig22}.

The difference of maximum and minimum bounds on the anomalous couplings
$\Delta\kappa_{V}$ and $\lambda_{V}$ (where $V=\gamma$, $Z$) can
be written as

\begin{equation}
\delta\Delta\kappa{}_{V}=\Delta\kappa_{V}^{\max}-\Delta\kappa_{V}^{\min},\:\delta\lambda_{V}=\lambda_{V}^{\max}-\lambda_{V}^{\min}\label{eq:eq3}
\end{equation}

The limits on anomalous couplings and the difference of maximum and
minimum bounds for electron beam energies $E_{e}=60$ and $140$ GeV
with integrated luminosities $L_{int}=10$ and $100$ fb$^{-1}$ at
LHeC with the unpolarized (polarized) electron beam are given in Tables
\ref{tab:tab3}-\ref{tab:tab6}.

\begin{table}[h]
\protect\caption{The limits on the anomalous couplings and the difference of maximum
and minimum bounds for electron beam energy of $E_{e}$=60 GeV with
$L_{int}$=10 fb$^{-1}$ for polarized and unpolarized electron beam.
\label{tab:tab3}}

\begin{tabular}{c|cccc|cccc}
\hline 
$P_{e}$ & $\Delta\kappa_{\gamma}$  & $\delta\Delta\kappa_{\gamma}$  & $\lambda_{\gamma}$  & $\delta\lambda_{\gamma}$ & $\Delta\kappa_{Z}$ & $\delta\Delta\kappa_{Z}$ & $\lambda_{Z}$ & $\delta\lambda_{Z}$ \tabularnewline
\hline 
-0.8 & {[}-0.366, 0.899{]}  & 1.265  & {[}-0.085, 0.148{]}  & 0.233 & {[}-0.152, 0.471{]} & 0.623 & {[}-0.016, 0.033{]} & 0.049 \tabularnewline
\hline 
0 & {[}-0.421, 0.940{]}  & 1.361  & {[}-0.094, 0.159{]}  & 0.253 & {[}-0.180, 0.498{]}  & 0.677  & {[}-0.018, 0.035{]} & 0.053 \tabularnewline
\hline 
0.8 & {[}-0.641, 1.177{]}  & 1.818  & {[}-0.141, 0.208{]}  & 0.349 & {[}-0.293, 0.611{]} & 0.904 & {[}-0.027, 0.044{]} & 0.071 \tabularnewline
\hline 
\end{tabular}
\end{table}

\begin{table}[h]
\protect\caption{Same as Table II but for $L_{int}$=100 fb$^{-1}$. \label{tab:tab4}}

\begin{tabular}{c|cccc|cccc}
\hline 
$P_{e}$ & $\Delta\kappa_{\gamma}$  & $\delta\Delta\kappa_{\gamma}$  & $\lambda_{\gamma}$  & $\delta\lambda_{\gamma}$ & $\Delta\kappa_{Z}$ & $\delta\Delta\kappa_{Z}$ & $\lambda_{Z}$ & $\delta\lambda_{Z}$ \tabularnewline
\hline 
-0.8 & {[}-0.237, 0.771{]}  & 1.008  & {[}-0.061, 0.124{]}  & 0.185 & {[}-0.088, 0.405{]} & 0.493 & {[}-0.011, 0.027{]} & 0.038 \tabularnewline
\hline 
0 & {[}-0.257, 0.777{]}  & 1.034  & {[}-0.064, 0.128{]}  & 0.192  & {[}-0.104, 0.412{]} & 0.516 & {[}-0.012, 0.028{]} & 0.040 \tabularnewline
\hline 
0.8 & {[}-0.356, 0.893{]}  & 1.249  & {[}-0.087, 0.153{]}  & 0.240  & {[}-0.147, 0.465{]} & 0.612 & {[}-0.016, 0.032{]} & 0.048 \tabularnewline
\hline 
\end{tabular}
\end{table}

\begin{table}[h]
\protect\caption{The limits on the anomalous couplings and the difference of maximum
and minimum bounds for electron beam energy of $E_{e}$=140 GeV with
$L_{int}$=10 fb$^{-1}$ for polarized and unpolarized electron beam.
\label{tab:tab5}}

\begin{tabular}{c|cccc|cccc}
\hline 
$P_{e}$ & $\Delta\kappa_{\gamma}$  & $\delta\Delta\kappa_{\gamma}$  & $\lambda_{\gamma}$  & $\delta\lambda_{\gamma}$ & $\Delta\kappa_{Z}$ & $\delta\Delta\kappa_{Z}$ & $\lambda_{Z}$ & $\delta\lambda_{Z}$ \tabularnewline
\hline 
-0.8 & {[}-0.255, 0.865{]}  & 1.120  & {[}-0.049, 0.088{]}  & 0.137 & {[}-0.208, 0.207{]} & 0.415 & {[}-0.003, 0.010{]} & 0.013 \tabularnewline
\hline 
0 & {[}-0.288, 0.895{]}  & 1.183  & {[}-0.052, 0.092{]}  & 0.144 & {[}-0.350, 0.170{]}  & 0.520  & {[}-0.005, 0.012{]} & 0.017 \tabularnewline
\hline 
0.8 & {[}-0.255, 1.035{]}  & 1.120  & {[}-0.070, 0.109{]}  & 0.179 & {[}-0.407, 0.373{]} & 0.780 & {[}-0.008, 0.014{]} & 0.022 \tabularnewline
\hline 
\end{tabular}
\end{table}

\begin{table}[h]
\protect\caption{Same as Table III but for $L_{int}$=100 fb$^{-1}$. \label{tab:tab6}}

\begin{tabular}{c|cccc|cccc}
\hline 
$P_{e}$  & $\Delta\kappa_{\gamma}$  & $\delta\Delta\kappa_{\gamma}$  & $\lambda_{\gamma}$  & $\delta\lambda_{\gamma}$ & $\Delta\kappa_{Z}$ & $\delta\Delta\kappa_{Z}$ & $\lambda_{Z}$ & $\delta\lambda_{Z}$ \tabularnewline
\hline 
-0.8 & {[}-0.182, 0.793{]}  & 0.975  & {[}-0.039, 0.079{]}  & 0.118 & {[}-0.143, 0.142{]} & 0.285 & {[}-0.001, 0.008{]} & 0.009 \tabularnewline
\hline 
0 & {[}-0.192, 0.798{]}  & 0.990  & {[}-0.041, 0.081{]}  & 0.122  & {[}-0.273, 0.089{]} & 0.362 & {[}-0.003, 0.009{]} & 0.012 \tabularnewline
\hline 
0.8 & {[}-0.251, 0.844{]}  & 1.095  & {[}-0.047, 0.086{]}  & 0.133  & {[}-0.253, 0.215{]} & 0.468 & {[}-0.004, 0.010{]} & 0.014 \tabularnewline
\hline 
\end{tabular}
\end{table}

\section{Conclusions}

The $WW\gamma$ and $WWZ$ anomalous interactions through the process
$ep\rightarrow$$\nu_{e}q\gamma X$ and $ep\rightarrow\nu_{e}qZX$
can be studied independently at the LHeC. We obtain two-parameter
accessible ranges of triple gauge boson anomalous couplings at LHeC
with the polarized and unpolarized beam at the energy of $E_{e}=60$
GeV and $E_{e}=140$ GeV. Our limits compare with the results from
two-parameter analysis given by ATLAS and CMS Collaborations \cite{ATLAS13,CMS13}.
We find that the LHeC with polarized electron beam improves the bounds
on anomalous couplings. The LHeC will give complementary information
about anomalous couplings compared to Tevatron and LHC. 
\begin{acknowledgments}
The work of O.C. is partially supported by State Planning Organisation
(DPT) - Ministry of Development under the grant No. DPT2006K-120470.
A.S. would like to thank Abant Izzet Baysal University Department
of Physics where of part this study was carried out for their hospitality.\end{acknowledgments}

\end{document}